\documentclass[
reprint,
comment,
superscriptaddress,
showpacs,
amsmath,
amssymb,
aps,
widetext,
floatfix,
]{revtex4-1}
\usepackage{graphicx}
\usepackage{dcolumn}
\usepackage{bm}
\usepackage{hyperref}
\usepackage{mathtools}

\usepackage[dvipsnames]{xcolor}

\begin{document}
\title{A Model of Densifying Collaboration Networks}%A Model of Heterogenous Densification}
\author
{Keith A. Burghardt}
\email{keithab@isi.edu}
\affiliation{Information Sciences Institute, University of Southern California, Marina del Rey, USA, 90292}%\\
%\normalsize{An Unknown Address, Wherever, ST 00000, USA}\\
% \normalsize{$^{2}$Department of Computer Science and Technology, Tsinghua University, Beijing, China, 100084}\\
%\normalsize{$^{2}$
\author{Allon G. Percus}
\affiliation{
Institute of Mathematical Sciences, Claremont Graduate University, Claremont, USA, 91711}%\\
%\normalsize{$^\ast$To whom correspondence should be addressed; E-mail:  keithab@isi.edu.}
%}
\author{Kristina Lerman}
\affiliation{Information Sciences Institute, University of Southern California, Marina del Rey, USA, 90292}

\date{\today}
\begin{abstract}
    Research collaborations provide the foundation for scientific advances, but we have only recently begun to understand how they form and grow on a global scale. Here we analyze a model of the growth of research collaboration networks to explain the empirical observations that the number of collaborations scales superlinearly with institution size, though at different rates (heterogeneous densification), the number of institutions grows as a power of the number of researchers (Heaps' law) and institution sizes approximate Zipf's law. This model has three mechanisms: (i) researchers are preferentially hired by large institutions, (ii) new institutions trigger more potential institutions, and (iii) researchers collaborate with friends-of-friends. We show agreement between these assumptions and empirical data, through analysis of co-authorship networks spanning two centuries. We then develop a theoretical understanding of this model, which reveals emergent heterogeneous scaling such that the number of collaborations between institutions scale with an institution's size. %We also demonstrate the model creates other realistic properties, such as strong modularity, a heavy-tailed degree distribution, clustering, and assortativity.
\end{abstract}
\maketitle

\section{Introduction}

%Research institutions, a category that includes universities, government labs, industrial labs, and national academies~\cite{hicks1996science,Taylor2019}, are critical to our understanding of scientific progress \cite{Raan2013,Deville2014,Way2019,Burghardt2020}. 
Science is largely a social endeavor. Research collaborations drive scientific discovery and produce more impactful work: papers with more co-authors garner more citations and appear in more prestigious venues~\cite{Wuchty2007,Dong2018}. Collaboration enables researchers to  mitigate the deleterious effects of the increasing complexity of knowledge~\cite{jones2009burden} by leveraging the diversity of expertise~\cite{page2019diversity} and different perspectives~\cite{yegros2015does}. Our understanding of the growing collaboration networks, however, is still in its infancy. A recent paper explored the role of research institutions in the growth of scientific  collaborations~\cite{Burghardt2020}, showing  that collaborations scale superlinearly with institution size: when an institution doubles in size, this creates roughly 30\% more collaborations per person.  
%Larger institutions therefore have a key advantage over smaller institutions. 
Crucially, the scaling laws are different for each institution; therefore, larger institutions typically receive more advantage from collaboration than others. Additionally, the paper showed that institutions vary in size by many orders of magnitude, with the distribution approximated by Zipf's law \cite{Zipf1949}, while the number of institutions scales sub-linearly with the number of researchers, in agreement with Heaps' law \cite{Lu2010,Simini2019}. The sublinear scaling implies that, even as more institutions appear, each institution gets larger on average, but this average belies an enormous variance.

Burghardt et al.~\cite{Burghardt2020} developed a stochastic model to explain these patterns, which we theoretically analyze here. In this model, a researcher appears at each time step and is preferentially hired by larger institutions (e.g., due to their prestige or funding). With a small probability, however, a researcher joins a newly appearing institution. The arrival of this new institution then triggers yet more new institutions to form in the future \cite{Tria2014}. Finally, once hired, researchers make connections to other researchers and their collaborators with an independent probability. Despite its simplicity, the model reproduces a range of empirical observations. 

The model combines three mechanisms. The first two mechanisms, known as Polya's Urn with Triggering, qualitatively reproduces the observed Heaps' law and Zipf's law \cite{Tria2014}. The mechanisms are the following (i) researchers are preferentially hired by large institutions, and (ii) new institutions trigger more potential institutions. The third mechanism is that researchers collaborate with friends-of-friends, which reproduces how collaborations scale with institution size \cite{Burghardt2020,Lambiotte2016,Bhat2016}. These model assumptions are tested here using bibliographic data from four fields: computer science, physics, math, and sociology. We show that the data are in broad agreement with model's assumptions about institution growth and links formation. We then explore the theory behind the model. We discover that the interaction of these mechanisms form novel emergent properties such as densification of links between emergent groups.  %While a parameter in our model
Finally, this model shows qualitative agreement with empirical statistics, such as significant community structure, heterogenous scaling laws of collaborations between institutions, and assortativity. 

The rest of the paper is organized as follows. First, we describe comparisons between bibliographic data patterns and assumptions of the model. Next, we show how network statistics qualitatively agree with expectations. Third, we develop a theoretical grounding for the model, and finally compare these theoretic predictions to simulations.

\section{A Model of Collaboration Growth}

Burghardt et al.~\cite{Burghardt2020} describe a stochastic growth model of institution formation that captures % elucidates
how institutions and collaborations jointly grow. They model the formation and growth of institutions by combining a P\'olya's urn-like model described by Tria et al., (\citeyear{Tria2014}), and a model of network densification \cite{Bhat2016,Lambiotte2016}. Unlike existing densification models~\cite{Leskovec2007,Bhat2016,Lambiotte2016}, Burghardt et al.'s model reproduces the heterogeneous densification of internal (within-institution) and external (between-institution) collaborations, and the non-trivial growth of institutions. In the Appendix, we show that realistic variants of this model will also produce qualitatively similar behavior.%as well as institution formation and growth with few parameters. 

\begin{figure}[t]
    \centering
    \includegraphics[width=1\columnwidth]{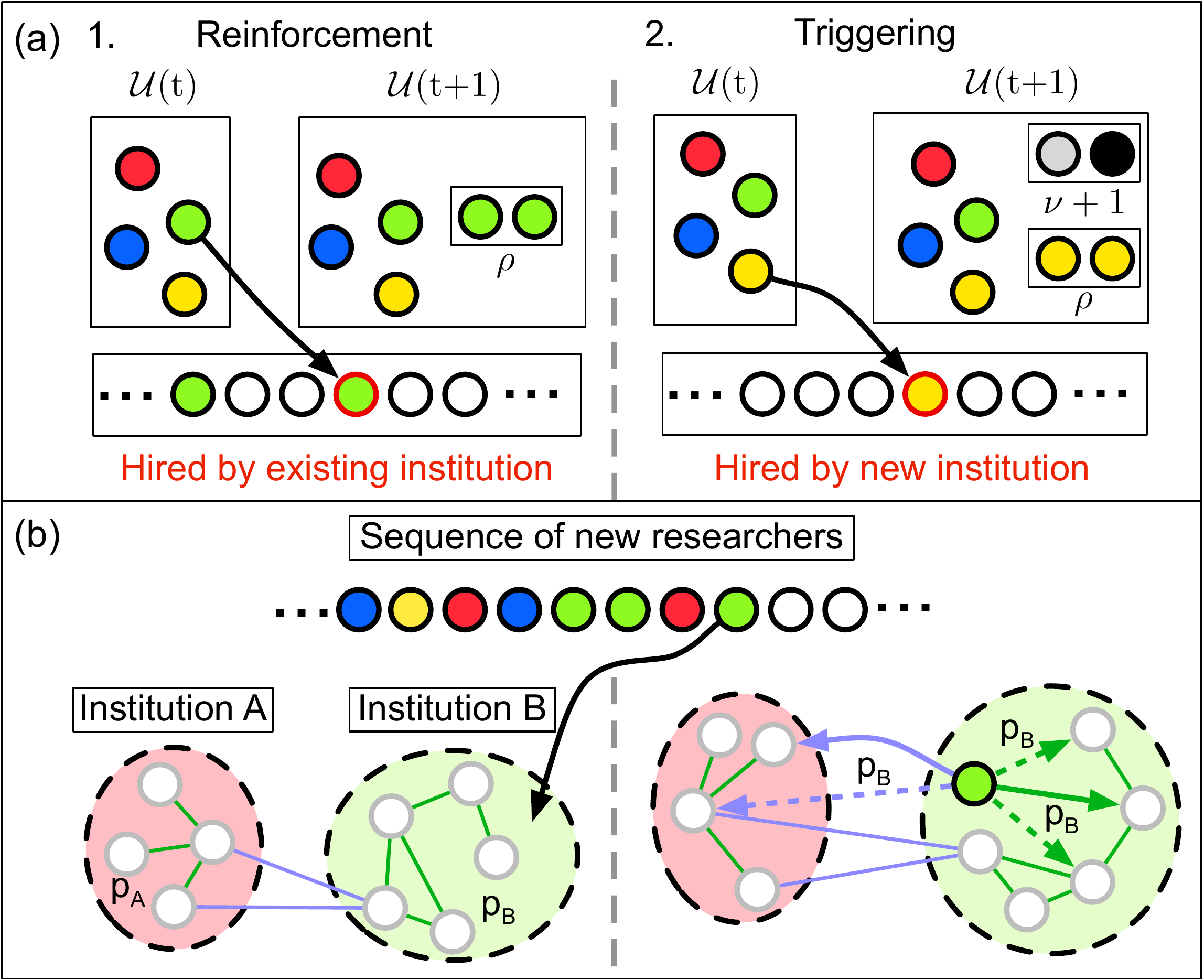}
    \caption{Schematic of our institution growth model. (a) New researchers are hired by an institution following a P\'olya's urn-like model \cite{Tria2014}. In this model, a new researcher is hired by an institution, denoted by a colored ball, picked uniformly at random from an urn. A new institution, where no researcher has been hired before, triggers $\nu+1$ new colors to enter the urn, increasing the likelihood of more new institutions to hire a researcher. Both new and old institutions experience reinforcement, where $\rho$ balls of the same color enter the urn. This creates a rich-get-richer effect where large institutions are more likely to hire a new researcher. (b) Each institution is composed of both internal collaborators (within each institution, green lines) and external collaborators (between institutions, purple lines). Once a researcher is hired, they choose one random internal and one random external collaborator. New collaborations are formed independently with probability $p_A$, if hired by institution A, and $p_B$ if hired by institution B. These new connections form triangles. Schematic taken from \cite{Burghardt2020}.
    }
    \label{fig:ModelSchematic}
\end{figure}

The model is as follows. Imagine an urn containing balls of different colors, with each color representing a different institution, as shown in  Fig.~\ref{fig:ModelSchematic}a. Balls are picked with replacement, each ball representing a newly hired researcher. The color of the picked ball is recorded in a sequence to denote  the institution that hires the researcher. After the ball is picked, $\rho$ new balls of the same color are added to the urn. This step, known as ``reinforcement'' (left panel of Fig.~\ref{fig:ModelSchematic}a)~\cite{Tria2014}, represents the additional resources and prestige given to a larger institution. If a ball with a new color that was not previously seen is picked, then $\nu+1$ uniquely-colored balls are placed into the urn. This step is known as ``triggering'' (right panel of Fig.~\ref{fig:ModelSchematic}a)~\cite{Tria2014}. The new colors represent institutions that are now able to form because of the existence of a new institution. This model predicts Heaps' law with scaling relation $\sim N^{\nu/\rho}$ and Zipf's law with scaling relation $\sim n^{-(1+\nu/\rho)}$ \cite{Tria2014}. In our simulations, we chose $\rho=4$ and $\nu=2$, which approximates scaling laws seen in data \cite{Burghardt2020}.%, which agrees well with the data shown in Fig.~\ref{fig:Scaling}. %\textbf{We also show in the SI (Supplementary Note 6: Analysis of Institution Growth Mechanism) that this model predicts the rate of institution growth is proportional to institution size (i.e.,\ follows a preferential attachment law), which we find is approximately correct. } 

\begin{figure}[tb!]
    \centering
    \includegraphics[width=\columnwidth]{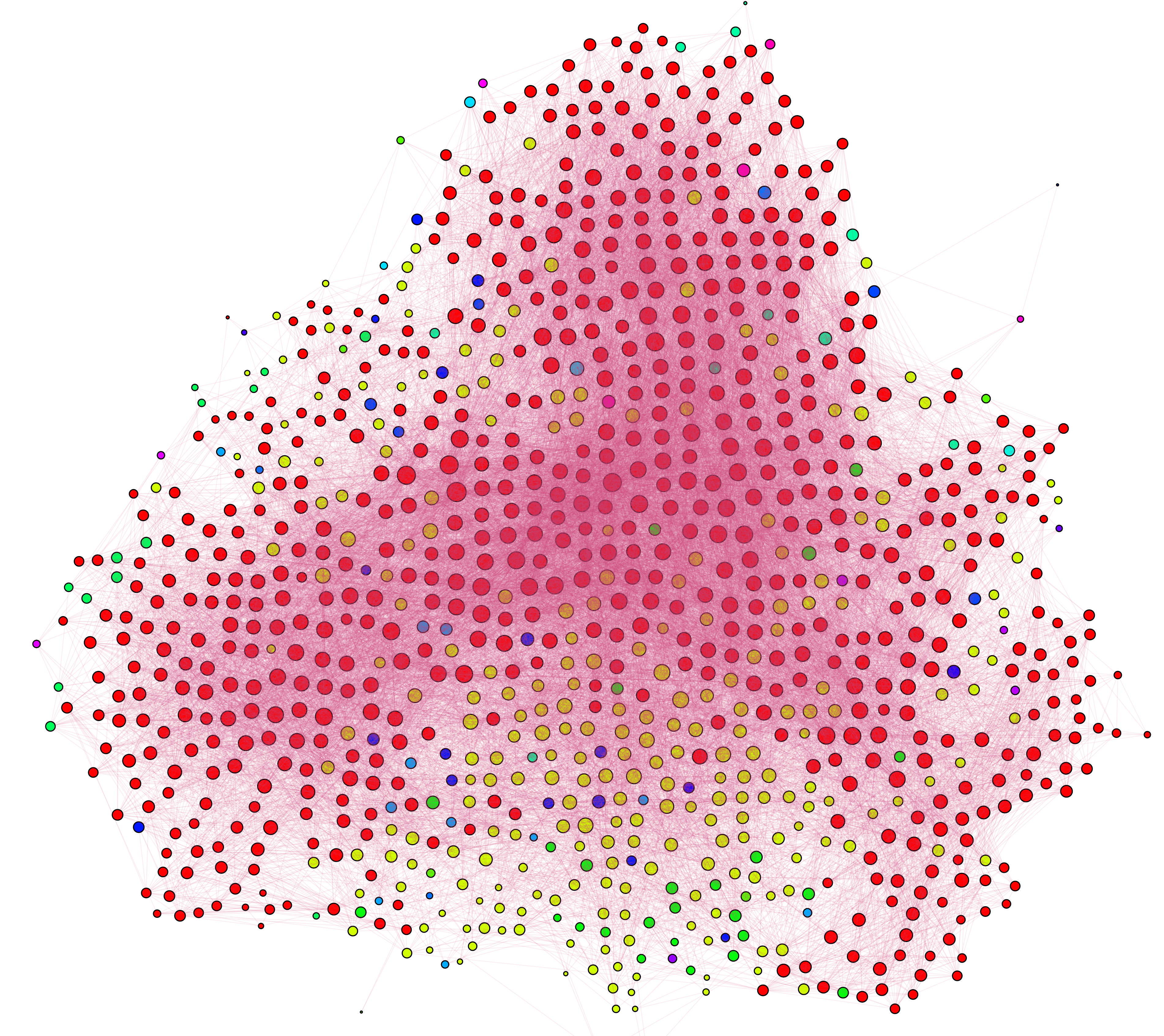}
    \caption{Example network from the simulation. Node sizes are proportional to degree and colors correspond to different institutions. Parameters are $\rho=4$, $\nu=2$, $\mu_p=0.6$, and $\sigma_p=0.1$ for $N=1000$ nodes.}
    \label{fig:netex}
\end{figure}

Next, we model heterogeneous and superlinear scaling of collaborations through a mechanism of network densification. Building on the work of  \cite{Lambiotte2016,Bhat2016}, each new researcher connects to a random researcher within the same institution, as well as an external researcher picked uniformly at random (left panel of Fig.~\ref{fig:ModelSchematic}b). Next, new collaborators are chosen independently from neighbors of neighbors with probability $p_{i}$, where $p_{i}$ is unique to each researcher's institution (right panel of Fig.~\ref{fig:ModelSchematic}b). We let $p_{i}$ be a Gaussian distributed random variable with mean, $\mu= 0.6$, and standard deviation,  $\sigma_\mu=0.1$ and truncated between 0 and 1. This parameter controls the heterogeneity we observe in collaboration scaling. An example output of this model is shown in Fig~\ref{fig:netex}. This plot demonstrates the complex patterns that appear in an otherwise simple model.

To summarize, the model has four parameters: $\rho$ and $\nu$, which control Zipf's law and Heaps' law, and $\mu_p$ and $\sigma_p$, which controls densification. In our analysis of the model, we fix $\rho=4$, $\nu=2$, $\mu_p=0.6$, which are in qualitative agreement with the statistics observed in empirical data \cite{Burghardt2020}.  While other plausible mechanisms for Zipf's law \cite{Gibrat1931,Eeckhout2004,Axtell2001}, Heaps' law  \cite{Simini2019}, or densification \cite{Leskovec2007} exist, the current model describes these patterns in a cohesive framework. %We will also show that the external scaling heterogeneity this model reproduces is a uniquely emergent property \cite{Burghardt2020}.

\section{Comparing Model Assumptions To Empirical Data}

We test the model against bibliographic data collected by Burghardt et al., (\citeyear{Burghardt2020}), based on data from the Microsoft Academic Graph~\cite{sinha2015overview,herrmannova2016analysis}. Author names and institutional affiliations have been extracted from when each paper was written allowing us to reconstruct the co-authorship network and institution size over time. The data Burghardt et al. analyze covers four different fields: computer science, physics, math, and sociology. They show that these results are robust to various assumptions of the data including whether institution size is defined as the cumulative number of authors affiliated with an institution, or in the other extreme, the number of affiliated authors who have written a paper in a particular year. For simplicity, we define institution size as the cumulative number of affiliated authors. Data parsing is described in greater detail in Burghardt et al.

\begin{figure}[tb!]
    \centering
    \includegraphics[width=0.33\textwidth]{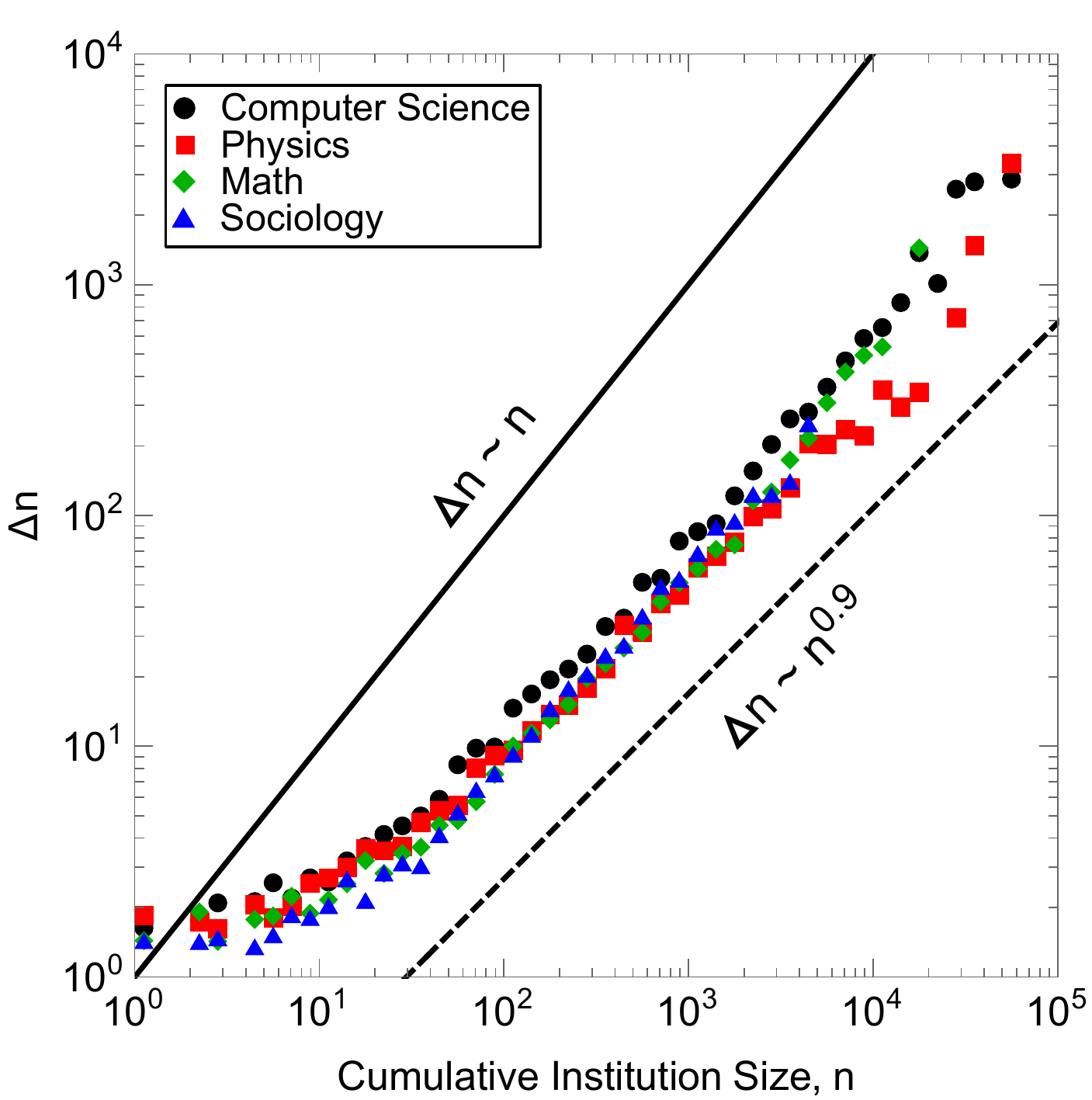}
    \caption{Rich-get-richer effect in institutions. The mean increase in institution size the next year as a function of its size, $n$, in the current year for the fields of computer science, physics, math, and sociology. The model predicts the rate of institution growth is proportional to its size (black line), and the best fit for data follows a power-law $\sim n^\alpha$ with $\alpha\approx 0.9$.
    }
    \label{fig:GrowthMechanism}
\end{figure}

This model predicts the rate of institution growth is proportional to institution size (i.e.,\ follows a preferential attachment mechanism), which we show is approximately correct in Fig.~\ref{fig:GrowthMechanism}. In the growth model, the probability an institution hires a researcher is proportional to the number of balls associated with that institution in an urn, and the number of balls is proportional to the institution's size. The probability an institution hires a researcher is therefore proportional to its size, $n$ (black line). When we compare to data, we see a slight deviation with growth proportional to $n^\alpha$ (dashed line) for $n>50$, where $\alpha$ is equal to $0.88\pm 0.02$, $0.80\pm 0.02$, $0.91\pm 0.02$, and $0.84\pm 0.01$ for computer science, physics, math, and sociology, respectively, based on linear regression. Alike to previous findings for preferential attachment \cite{Jeong2003,Sheridan2018}, this figure demonstrates that the mechanism approximately captures the relationship between size and growth.

\begin{figure}[tb!]
    \centering
    \includegraphics[width=0.33\textwidth]{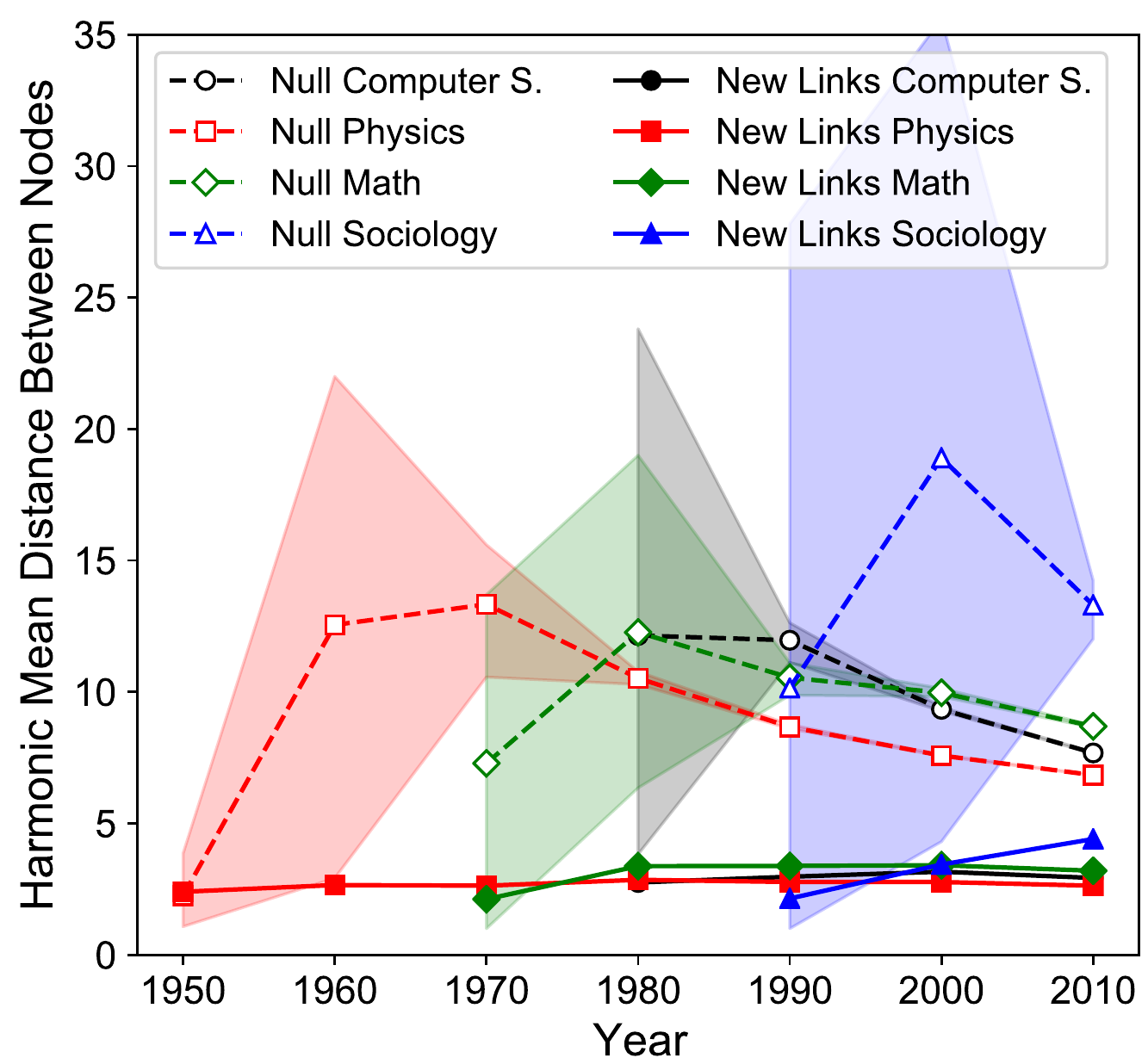}
    \caption{New links form between local nodes. For each institution we compare the geometric mean distance between nodes just before a link forms to random nodes (null model). Shaded ares are 95\% confidence intervals.
    }
    \label{fig:Densify}
\end{figure}

Next, the model assumes new connections are formed locally in order for networks to densify \cite{Burghardt2020,Lambiotte2016,Bhat2016}. We test this in Fig.~\ref{fig:Densify}, in which we compare the geodesic distance between researchers before they form new collaborations (solid markers) with the distance between random researchers (null model, open markers). The model would predict that collaborations form between researchers who are two collaborations from each other. For example, if one researcher was Paul Erd\"os, then the other researcher would have had an Erd\"os number of two prior to collaborating \cite{Castro1999}.

In this figure, new collaborations are defined as those that appear the next year and never appeared in any previous year, and plots were made for data 10 years apart. For example, new links were those that first appeared in 1951, 1961, 1971, etc. We take the harmonic mean of the geodesic distance to account for uncommon cases in which components are disconnected, and therefore the geodesic distance is infinity. To determine error bars (shaded regions) in the null model, we use a form of bootstrapping. We repeat the following step $M$ times: we find the mean distance of $m$ random researchers, where $m$ is the number of new links formed the next year. We let $M$ be $100$ for computer science and physics, and $300$ for math and sociology. Error bars are simply the 95\% quantiles of these bootstrapped data. Due to the cost of finding geodesic distances, computing these null model error bars took roughly 50 computer-hours to complete on 3.7 GHz Intel Core i5 processors. Comparing the distances of new research collaborations to this null model, we observe that researchers collaborate locally, often with high statistical significance. In rough agreement with the model, we see that researchers connect to one another when they are two to three collaborations apart, on average. 

\begin{figure}[thb!]
    \centering
    \includegraphics[width=\linewidth]{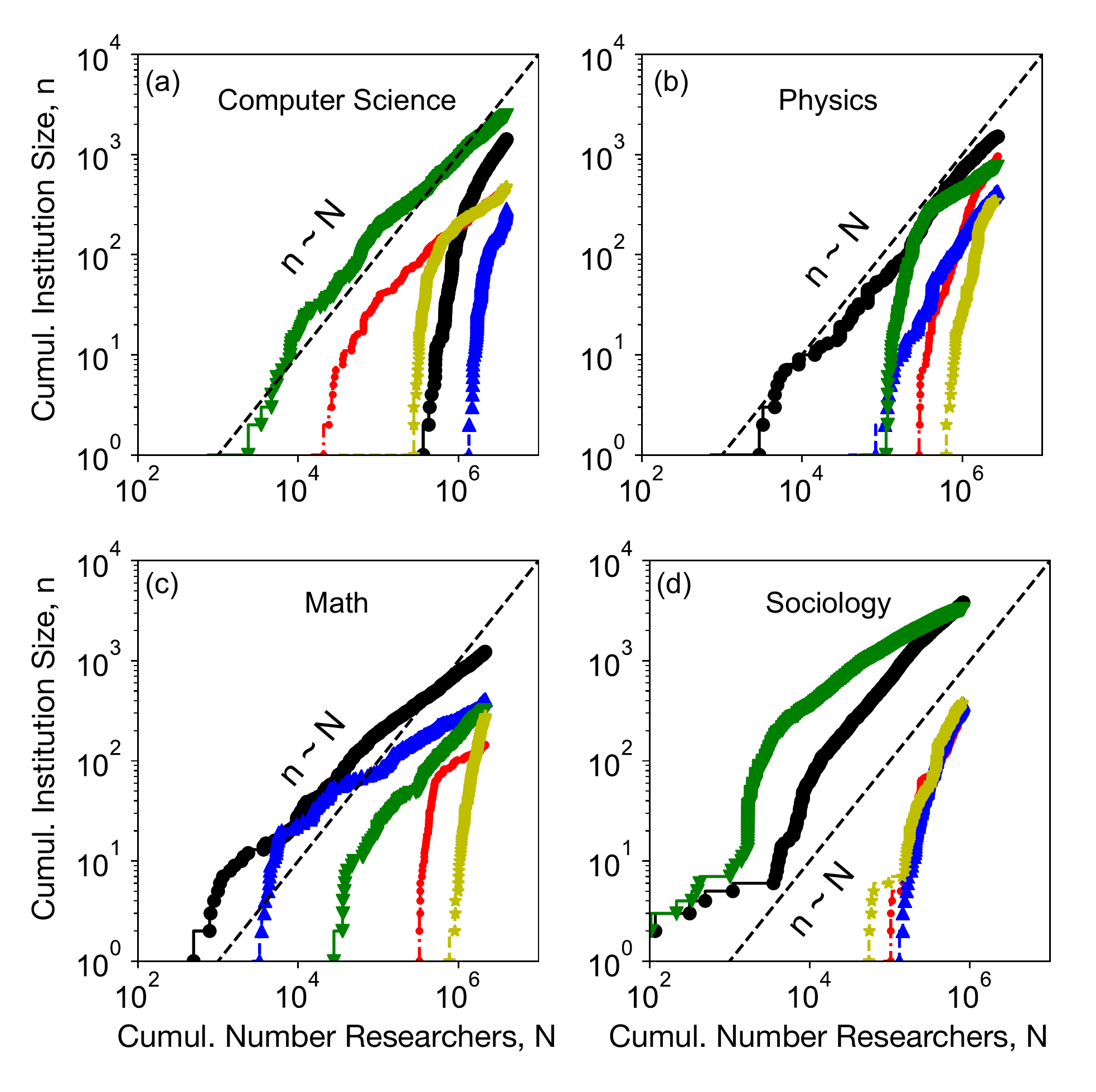}
    \caption{Cumulative institution size, $n$, versus total number of researchers hired, $N$. The model predicts $N\sim n$ (dashed lines). Five example institutions are shown for each field, (a) computer science, (b) physics, (c) math, and (d) sociology.
    }
    \label{fig:InstSizeVsHires}
\end{figure}
Finally, the first step of the model, institution growth, has the same set of mechanisms as Tria et al., (\citeyear{Tria2014}). Supplementary material of \cite{Tria2014}~Eq. 4--5 implies that institutions should grow proportional to ``time'' in the model. Time in this case is the cumulative number of researchers hired within any university, $N$, therefore the size of the institution, $n$, should be proportional to $N$. For example, if there are fifty institutions that have hired 1,000 researchers in total, then once 2,000 researchers have been hired, the number of researchers within each institution should approximately double (assuming that the number of new institutions that appear is small). We test this qualitatively in Fig.~\ref{fig:InstSizeVsHires} for each field. We find that the initial growth is usually much faster than linear (dashed line), and sometimes the asymptotic growth rate is sub-linear (e.g., the largest institution in Fig.~\ref{fig:InstSizeVsHires}d). That said, we also often see approximately linear growth. Overall, these results give mixed support for the hypothesis on average, but the variations from linear growth suggest the model, perhaps because of its simplicity, does not fully capture the data.

\section{Qualitative Statistics}

\begin{figure}[tb]
    \centering
    \includegraphics[width=0.5\textwidth]{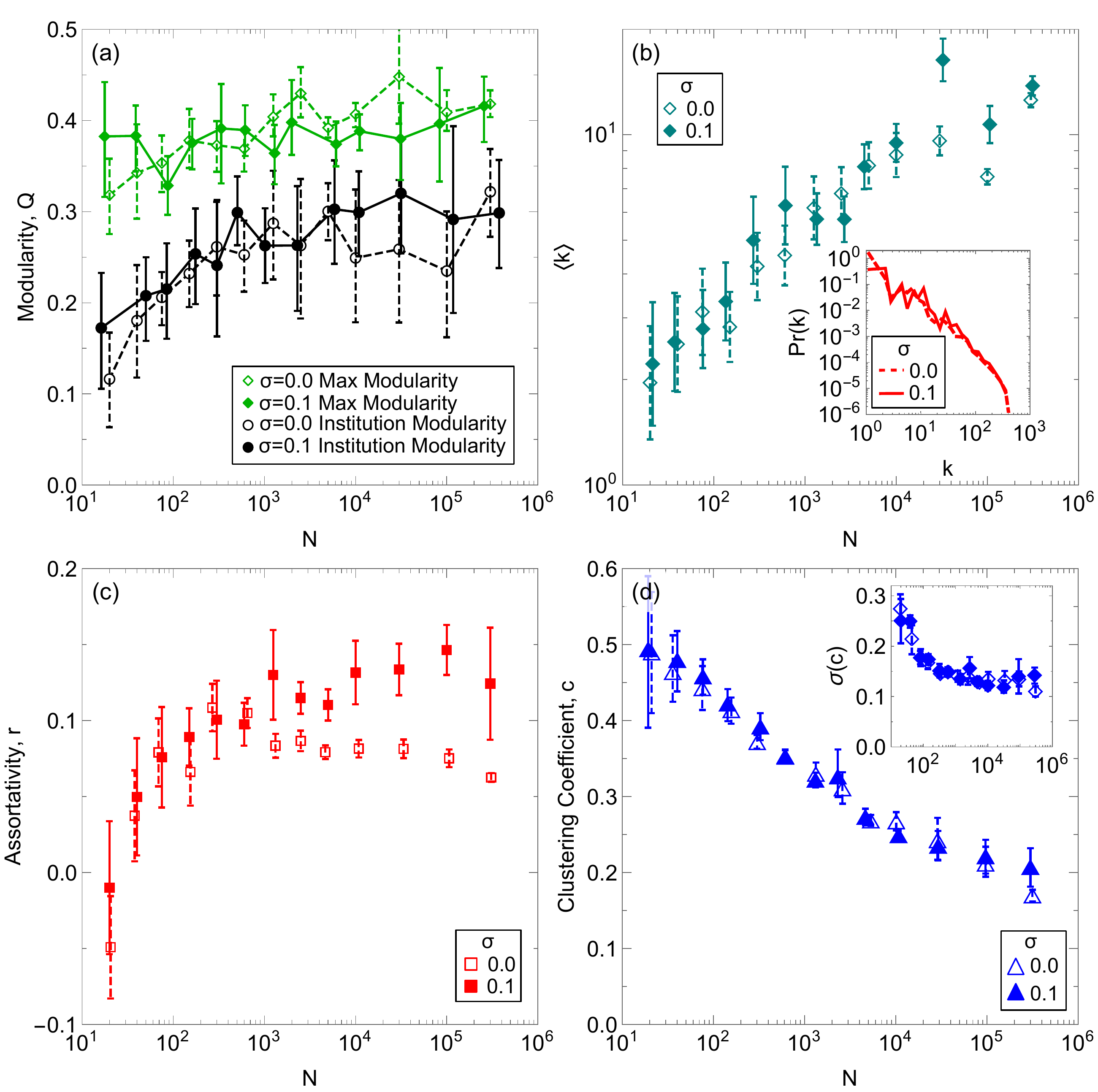}
    \caption{Network statistics vesus N. (a) Modularity based on a greedy modularity maximizing method \cite{Clauset2004}, and similar modularity values with communities defined as institutions, with $\sigma_p=0.0$ and 0.1 (all values are larger than expected by chance \cite{Guimera2004}). (b) Network degree versus N. Inset: degree distribution for $\sigma_p=0.0$ and 0.1. (c) Assortativity versus N, showing consistently higher values for $\sigma_p=0.1$ than $\sigma_p=0.0$. (d) Clustering coefficient decreases approximately logarithmically with N.
    }
    \label{fig:SimStats}
\end{figure}

Next, we measure network statistics of model simulations to check whether these statistics are realistic. We also find that the community structure, densification, assortativity, and clustering, shown in Fig.~\ref{fig:SimStats}, are comparable to real networks. First, this model naturally produces community structure if we define ``communities'' as institutions. The modularity of these communities is nearly as high as that from a greedy modularity maximization method \cite{Clauset2004}, possibly because the institution-based communities are alike to the stochastic block model \cite{Abbe2017}. In the stochastic block model, communities are defined as a collection of nodes that are more likely to connect to each other than to outside nodes. Similarly, institutions in the model have different probabilities of forming connections within and between other institutions.

Second, we find that degree increases with $N$ as a power law, known as network densification \cite{Leskovec2007,Lambiotte2016} (Fig.~\ref{fig:SimStats}b). While we designed individual institutions to densify, these results still reproduce previous global analysis demonstrating overall densification of the network. In addition, we see in the inset that the degree distribution is heavy-tailed, much like real networks \cite{Barabasi1999}. Importantly, the model has no explicit degree preferential attachment mechanism; this distribution is an emergent property. Dependence of the degree distribution with $N$ can be seen in the Appendix Fig.~\ref{fig:DegreeDist}.

Next, we find that assortativity increases with $N$ and is comparable to real social networks, including co-authorship networks we aim to model \cite{Newman2003b} (Fig.~\ref{fig:SimStats}c). Interestingly, however, assortativity begins to decrease again if $\sigma=0.0$. When $\sigma>0.0$, this model reproduces the heterogenous densification, seen in empirical data \cite{Burghardt2020}, as well as consistent positive values of assortativity. Finally, the local clustering coefficient decreases logarithmically, as shown in Fig.~\ref{fig:SimStats}d. In contrast, a random network has a clustering coefficient that decreases as $1/n$ \cite{Newman2003}. The model's clustering coefficient is comparable to real data of a variety of sizes \cite{Boccaletti2006}. Whether clustering coefficient is stable in real data \cite{Ostroumova2013} or decreases with $n$ should be explored in the future. The variance of the local clustering coefficient within each network (inset of Fig.~\ref{fig:SimStats}d), is also wide and should be compared to empirical data in the future. %Comparison with real data should also be explored in the future.

%We show correlations between internal and external collaboration exponents in cross-section data and longitudinal data. In Fig.~\ref{fig:CrossInstituteCollabCorrel}, we see a strong correlation between the internal and external collaboration exponents in cross-sectional data. %An example of the best fit lines can be seen in Fig.~2a--b of the main text. 
% Each field has Spearman rank correlations with $s>0.88$ ($P<10^{-6}$), suggesting strong correlations between these two exponents. This is in contrast to simulations, where among 10 simulations, the average value of s is -0.31 (range is between -0.83 and 0.28, $P<10^{-6}$). An example of a plot of internal and external collaboration exponents is seen in Fig.~\ref{fig:SimExCollabCorrel}. That said, both simulations and data show strong differences between cross-sectional and longitudinal data exponents.

\begin{figure}[tb!]
    \centering
    \includegraphics[width=0.3\textwidth]{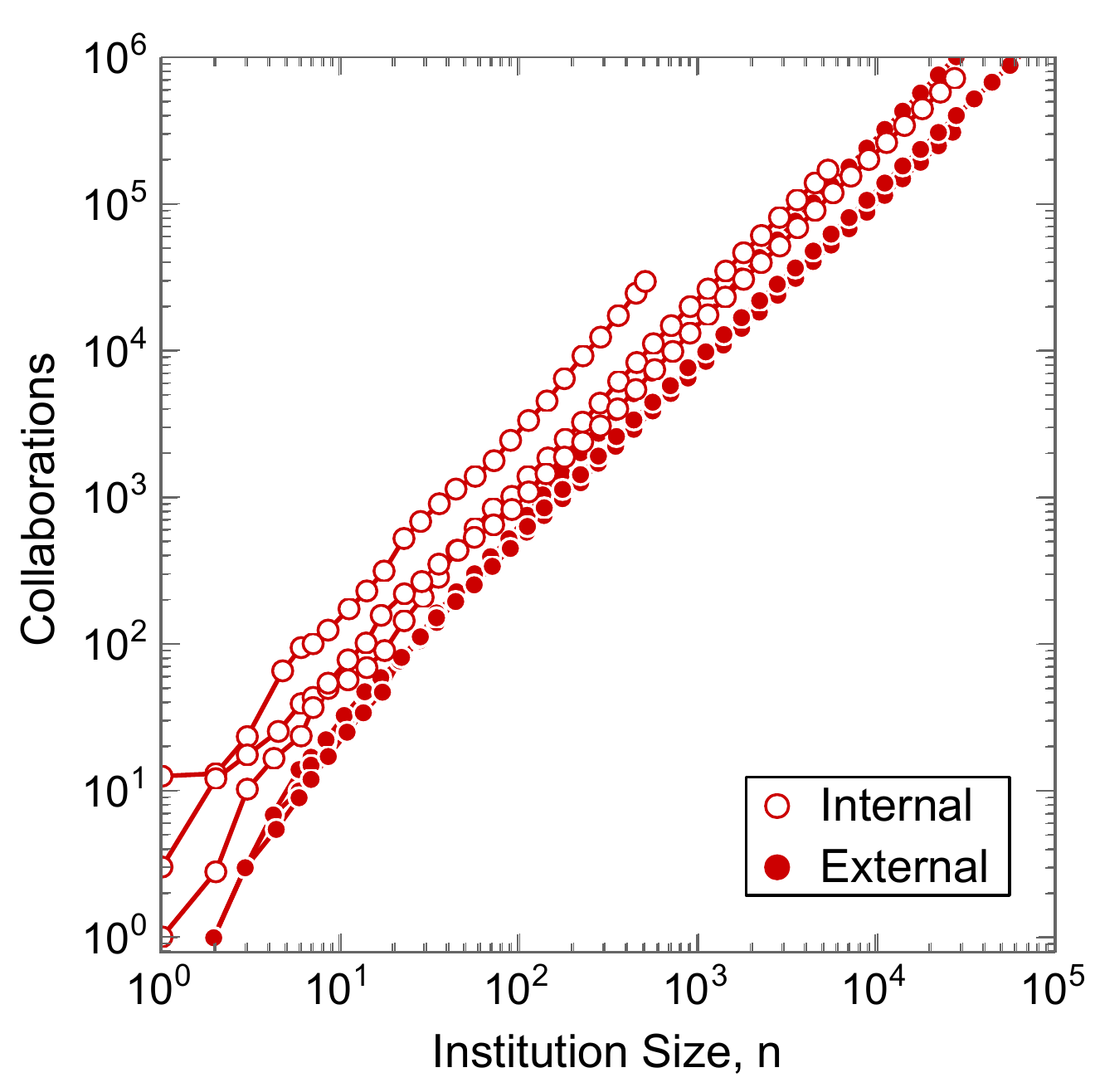}
    \caption{Simulations of the number of collaborations versus institution size. Internal scaling (within an institution) and external scaling (between institutions) is superlinear and varies between institutions.
    }
    \label{fig:SimScaling}
\end{figure}

\section{Analysis of the Model} %Simulation Theory}

Next, we develop a theoretical understanding of the model. 
%How are the scaling laws for internal and external collaborators in the model expected to relate to institution size? And why are the scaling laws poorly correlated with each other? We begin with the 
We first analyze the scaling properties of internal collaborations. Because the mechanism to form internal collaborations ignores all nodes and links besides those within the institution itself, we can consider the institution's internal collaboration network as an isolated network. The mechanism to make collaborations within this network can therefore be reduced to that of a previous model \cite{Bhat2016,Lambiotte2016}. The number of internal collaborations, $L_{\text{int}}$ increases with institution size, $n$ via the following formula
%\begin{equation}
\begin{align}
     L_{\text{int}}(n+1) &= L_{\text{int}}(n) + 1 + p \langle k_{\text{int}}\rangle \\
     &=L_{\text{int}}(n) + 1 + 2 p L_{\text{int}}(n)/n
     \end{align}
 %\end{equation}
     where $\langle k_{\text{int}}\rangle$ is the mean number of internal collaborations per researcher, equal to $2  L_{\text{int}}(n)/n$. Intuitively, we add an edge by default, plus $p \langle k_{\text{int}}\rangle$ edges through additional collaborations. Using the results from previous papers \cite{Bhat2016,Lambiotte2016}, we find that 
     \begin{equation}\label{eq:intscale}
         L_{\text{int}}(n) = \begin{cases}
         \frac{n}{1-2 p} & p<1/2\\
         n\text{ln}(n) & p = 1/2\\
          A(p) n^{2 p} & p > 1/2
         \end{cases}
     \end{equation}
     where $A(p) = [(2 p-1)\Gamma(1+2 p)]^{-1}$. The scaling constants and exponents in this theory are taken across all realizations. In practice, however, the exponent works well for large institutes, and underestimates the exponent for small institutes, most likely because of finite size effects. 
    
\begin{figure}
    \centering     \includegraphics[width=0.35\textwidth]{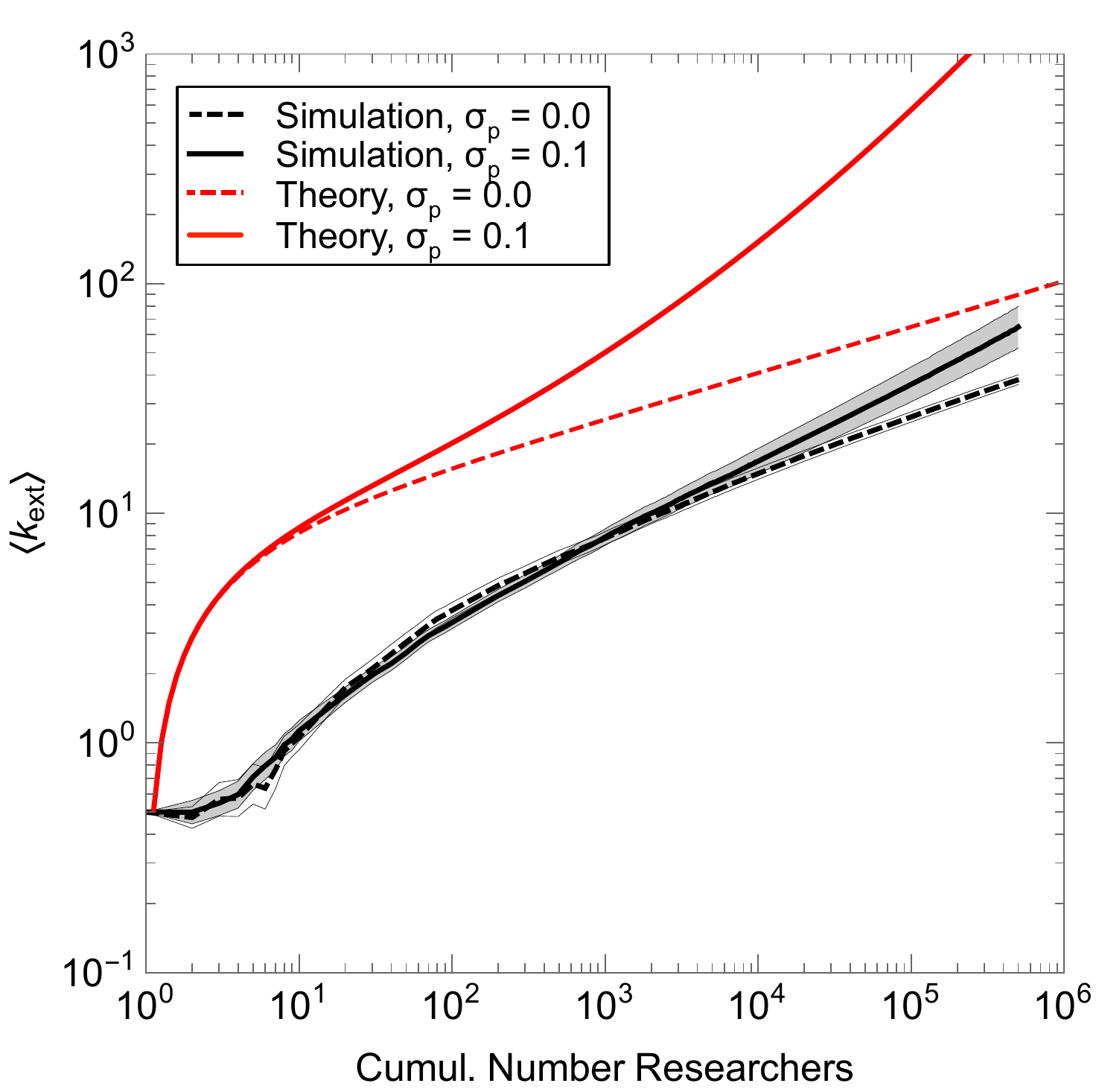}
     \caption{Mean degree of external collaborations, i.e., with researchers at different institutions, versus the cumulative number of researchers for several simulations. Solid black line are simulations with $\sigma_p=0.1$, dashed black line are simulations with $\sigma_p=0.0$. Solid red line is finite $\sigma_p=0.1$ theory, and dashed red line is $\sigma_p=0.0$ theory, shown in Eq.~\ref{eqn:Nkext}. Shaded ares are 95\% confidence intervals of the mean.}
     \label{fig:KMean}
 \end{figure}

     External scaling laws are much more nuanced, and require significant amounts of new analysis. We have two goals in our analysis. First, we want to show that internal and external collaboration exponents are superlinear. Second, we want to understand why internal and external collaboration exponents are poorly correlated. To this end, we start with a similar equation as before, but this time for external collaborations, $L_{\text{ext}}$:
     \begin{equation}\label{eqn:Lext}
     L_{\text{ext}}(n+1) = L_{\text{ext}}(n) + 1 + p \langle k_{\text{ext}}\rangle
     \end{equation}
    Our goal is to first find $\langle k_{\text{ext}}\rangle$, mean number of internal collaborations per researcher at external institutions. This value is surprisingly non-trivial compared to internal collaborations. First, we note that the first researcher is chosen at random among all researchers, meaning there is a preference to attach to researchers in larger institutes. While the institution size follows Zipf's law \cite{Tria2014}, 
     \begin{equation}
         p(n) = \frac{\nu}{\rho} n^{-(1+\nu/\rho)},
    \end{equation}
    where we take the discrete size $n$ to be continuous, which works well for large institution sizes. The preference to attach to larger institutes means that we choose an institute of size $n_{ext}$ with probability
       \begin{equation}
         q(n) = \frac{n p(n)}{\langle n\rangle}
     \end{equation}
    where
     \begin{align}
     \langle n\rangle &= \int_1^N dn~  n \left( \frac{\nu}{\rho} n^{-(1+\nu/\rho)}\right)\\
     &\sim \frac{\nu}{\rho-\nu} N^{1-\nu/\rho}
    \end{align}
     Because $\nu/\rho < 1$, we discover that $\langle n\rangle$ diverges. Therefore, we set of cut-off equal to the total number of researchers, $N$. In full form, $q(n)$ is:
 \begin{equation}
         q(n) = \frac{(\rho-\nu) n^{-\nu/\rho}}{\rho N^{1-\nu/\rho}}
     \end{equation}
     Moreover, by construct, we have the probability of $p$, $f(p)$, be Gaussian distributed, with mean $\mu_p$ and variance $\sigma_p^2$. Finally, $k_{\text{ext}}$ for an arbitrary institution is $2 L_{\text{int}}(n)/n$. Putting all this together, we discover that
   
       \begin{widetext}
  \begin{equation}\label{eqn:noapprox}
 %\begin{align}
     %\begin{strip}
         \langle k_{\text{ext}}\rangle =\frac{2 (\rho-\nu)}{\sigma\sqrt{2\pi} \rho N^{1-\nu/\rho}} \int_1^N dn~ \left\{n^{-\nu/\rho} \int_0^{1/2} dp~\frac{\text{exp}[-(p-\mu_p)^2/(2\sigma_p^2)]}{1-2p}+ \int_{1/2}^{1}dp~ \frac{n^{2p-1-\nu/\rho}\text{exp}[-(p-\mu_p)^2/(2\sigma_p^2)]}{(2p-1)\Gamma(1+2p)}\right\}
     %\end{strip}
    \end{equation}
      \end{widetext}
 %\end{align}
     Sadly, this equation is not simple to solve. First, it diverges near $p=1/2$. At this special point, the scaling law approaches $L_{\text{int}}(n)\sim n\text{ln}(n)$, which is why the assumptions around $p\simeq 1/2$ break down. If we take the ends of the integrals to be 0 to $1/2-\epsilon$ and $1/2+\epsilon$ to 1, then $\langle k_{\text{ext}}\rangle$ becomes a constant proportional to $\text{ln}(1/\epsilon)$. If this value is small compared to $N$, however, then from Eq.~\ref{eqn:Lext}, $L_{\text{ext}}(n)\sim n$, which does not agree with our findings. On the other hand, $\langle k_{\text{ext}}\rangle$ (and therefore $\text{ln}(1/\epsilon)$) cannot be larger than $N-1$. In other words, we can only connect to as many as nodes as there are in the network. If we assume $\langle k_{\text{ext}}\rangle\sim N$, then from Eq.~\ref{eqn:Lext}, $L_{\text{ext}}(n)\sim n^2$. This demonstrates a breakdown in the assumptions of a naive approximation of Eq.~\ref{eqn:noapprox}. 
     
     That being said, we can make perturbative expansions around $\mu_p$ assuming $\sigma_p$ is small. In this limit, $\text{exp}[-(p-\mu_p)^2/(2\sigma_p^2)]$ approaches zero faster than $1/(2 p-1)$ approaches infinity, therefore we can integrate around $\mu_p$.
         If $\sigma_p$ is small, we can focus on $p>1/2$ (assuming $\mu_p>1/2$) and note that $\text{exp}[-(p-\mu_p)^2/(2\sigma_p^2)]$ varies much more than the denominator, which we can approximate as $(2\mu_p-1)\Gamma(1+2\mu_p)$. On the other hand, because $n$ is assumed to be large, a small variation in $p$ could significantly change the numerator, therefore $n^{2 p}$ is not approximately $n^{2 \mu_p}$ unless $\sigma\rightarrow 0$, thus the Gaussian distribution becomes a Dirac delta function. In the small $\sigma_p$ limit,
     \begin{widetext}
     \begin{equation}
         \langle k_{\text{ext}}\rangle = \frac{2 (\rho-\nu)}{\sigma_p \sqrt{2 \pi}\rho N^{1-\nu/\rho}}\int_1^N dn~n^{-(1+\nu/\rho)}\int_{1/2}^1 dp~ \frac{\text{exp}[2 p\text{ln}(n)-(p-\mu_p)^2/(2 \sigma_p^2)]}{(2 \mu_p-1)\Gamma(1+2\mu_p)}
     \end{equation}
     \end{widetext}
     because the PDF quickly approaches 0 around $p=\mu_p$, we can extend the integral of $p$ to $\pm \infty$. Once we integrate, the result becomes
      \begin{widetext}
     \begin{equation}
         \langle k_{\text{ext}}\rangle = \frac{2 (\rho-\nu)}{(2 \mu_p-1)\Gamma(1+2\mu_p)\rho N^{1-\nu/\rho}}\int_1^N dn~n^{2\mu_p-(1+\nu/\rho)} \text{exp}[2 \sigma_p^2\text{ln}(n)^2]
     \end{equation}
     \end{widetext}
     after integrating over $dn$, the result become\newpage
      \begin{widetext}
      \begin{equation}
      %\begin{aligned}
    \label{eqn:kextexact}
     \langle k_{\text{ext}}\rangle =\frac{2 (\rho-\nu)}{\sqrt{2}\sigma(2 \mu_p-1)\Gamma(1+2\mu_p)\rho N^{1-\nu/\rho}} \left\{ F\left[\frac{\nu -2 \mu_p \rho }{2 \sqrt{2} \rho  \sigma_p }\right]+N^{-\frac{\nu }{\rho}+2 \mu_p+2 \sigma_p^2 \log (N)} F\left[\frac{4 \rho  \log (N) \sigma_p^2-\nu +2 \mu \rho }{2 \sqrt{2} \rho  \sigma_p}\right]\right\},
     %\end{aligned}
     \end{equation}
       \end{widetext}
    where $F$ is the Dawson function \cite{Abramowitz1972}. If, on the other hand, $\sigma_p$ is zero, then we replace the Gaussian distribution with a Dirac delta and the equation becomes
    \begin{widetext}
    \begin{equation}
         \langle k_{\text{ext}}\rangle_{\sigma_p=0} = \frac{2 (\rho-\nu)}{(2 \mu_p-1)\Gamma(1+2\mu_p)\rho N^{1-\nu/\rho}}\int_1^N dn~n^{2\mu_p-(1+\nu/\rho)}
    \end{equation}
    \end{widetext}
        \begin{widetext}
    \begin{equation}
        \langle k_{\text{ext}}\rangle_{\sigma_p=0} = \frac{2 (\rho-\nu)}{(2 \mu_p-1)\Gamma(1+2\mu_p)\rho  (2\mu_p-\nu/\rho)}(N^{2\mu_p-1}-1)
     \end{equation}
         \end{widetext}
     We compare this to simulation data in Fig.~\ref{fig:KMean}, and find similar scaling behavior, although the values are off by a factor of 10, possibly due to the finite size of most institutions, where the scaling laws assumed above might not hold. To understand the long-term behavior, however, we can take the limit that $N\rightarrow\infty$
     \begin{equation}\label{eqn:Nkext}
     \langle k_{\text{ext}}\rangle \approx 
        \begin{cases}
         C_1 N^{2\mu_p-1} & \sigma_p = 0~(p = \mu_p)\\
         C_2 \frac{N^{2\mu_p-1+2 \sigma_p^2 \text{ln}(N)}}{\text{ln}(N)} & \sigma_p\ll1
        \end{cases}
    \end{equation}
     where 
     \begin{equation}
     C_1 = \frac{2(\rho-\nu)}{(2\mu_p-\nu/\rho)(2\mu_p-1)\Gamma(2\mu_p+1)}
     \end{equation}
     and 
     \begin{equation}
     C_2=\frac{\rho-\nu}{\rho \sigma_p^2(2-\mu_p-1)\Gamma(2\mu_p+1)}
     \end{equation}
     We notice that variance \emph{increases} the mean degree, but also that that, for finite $\sigma_p$, the scaling relation is not a power law. %That being said, we take the limit $\sigma\ll \mu$ in order to make analytic conclusions, which does not appear appropriate when our data has $\mu=0.6$ and $\sigma=0.1$, therefore this formula must break down when we are far outside that limit. That being said, we can make a simpler model that gives intuition about the scaling laws of $\langle k_{ext}\rangle$, as shown in Fig.~\ref{fig:SimpleSchematic}. First, instead of a Gaussian distribution of $p$, we instead  assume $p$ is a top hat distribution, with range $2\eta$ and mean $\mu$ (Fig.~\ref{fig:SimpleSchematic}a). Next, an institution's degree, $k_{ext}$ is, to first order, $\sim 1$ if $p<1/2$, and $\sim N^{2 p-1}$ if $p>1/2$. We ignore the coefficients, $A(p)$, that might affect degree (Fig.~\ref{fig:SimpleSchematic}b). Following the same procedure as above, the mean degree, $\langle k_{ext}\rangle$ is:
     What we are interested in, however, is how $\langle k_{\text{ext}}\rangle$ depends on $n$, the institution size. Previous research shows, to first order, that $n = N/N_i$, where $N_i$ is the number of researchers when the first institute formed (c.f., Supplementary materials Eq. 4 of \cite{Tria2014}). Substituting this into Eq.~\ref{eqn:Nkext}, we get $\langle k_{\text{ext}}\rangle$ as a function of $n$. %we find that, while the constant depends on when the institute formed, the asymptotic equation is of the form:
%     %\begin{equation}\label{eqn:Nkext}
%     %\langle k_{ext}\rangle = 
%     %    \begin{cases}
%     %  \sim\frac{n^{2\mu-1+2 \sigma^2 \text{ln}(n)}}{\text{ln}(n)} & \sigma>0\\
%     %    \sim n^{2\mu-1} & \sigma = 0~(p = \mu)
%     %    \end{cases}
%     %\end{equation}
  We can finally substitute $\langle k_{\text{ext}}(n)\rangle$ into Eq.~\ref{eqn:Lext}, and notice that $\langle k_{\text{ext}}\rangle$ does not depend on $L_{\text{ext}}$, in contrast to internal collaborations. Knowing that $L_{\text{ext}}(1) = 0$, this iterative equation can be solved in the form of a series:
%   %\begin{equation}
    \begin{align}
        L_{\text{ext}}(n) &= \sum_{j=1}^{n-1} 1 + p\langle k_{\text{ext}}\rangle(j)\\
         &=n-1 + p \sum_{j=1}^{n-1} \langle k_{\text{ext}}\rangle (j)
         \end{align}
     %\end{equation}
     sadly, there is in general no simple formula for this series, although if $\sigma_p=0$
    %because $n$ is small compared to the asymptotic formula,
    \begin{equation}\label{eqn:Lext0}
     L_{\text{ext}}(n)_{\sigma_p=0} \sim 
         %\begin{cases}
       %p \sum_{j=1}^{n-1} j^{2\mu-1+2\sigma^2\text{ln}(j)}/\text{ln}(j) & \sigma>0\\
         p \sum_{j=1}^{n-1} j^{2\mu_p-1}=H(n-1,1-2\mu_p)% & \sigma = 0~ (p = \mu)
         %\end{cases}
     \end{equation}

     \begin{figure}
     \centering
     \includegraphics[width=0.5\textwidth]{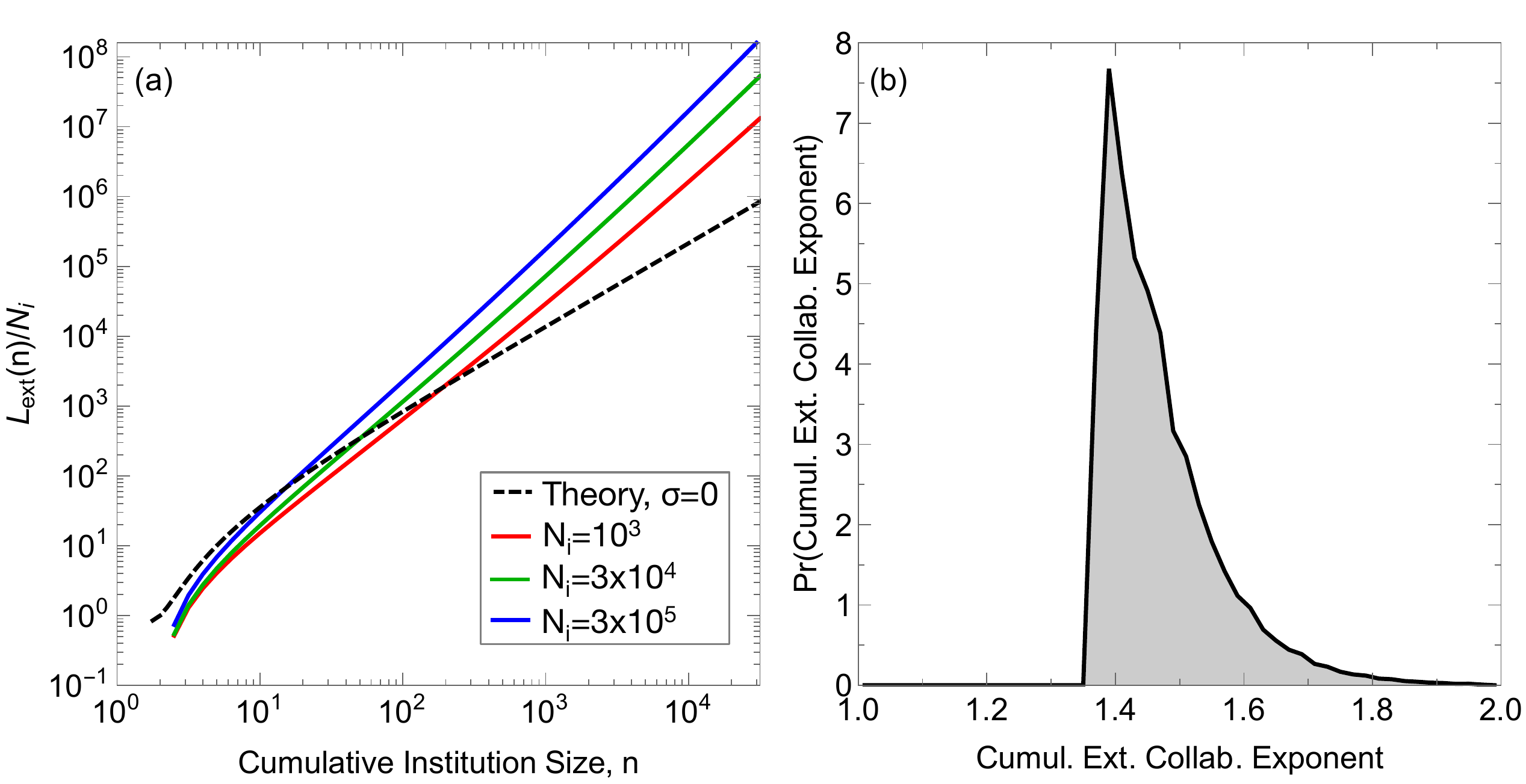}
     \caption{Theoretical scaling laws. (a) External degree, $L_{\text{ext}}$, normalized by the cumulative number of researchers when the institution first formed, $N_i$, versus the institution size. Solid lines are theory with $\sigma_p>0$ and various values of $N_i$. Dashed line is Eq.~\ref{eqn:Lext0}. (b) A theoretical histogram of scaling exponents, fit for $n>10$, after $N=10^6$. 
     }
     \label{fig:LextTheory}
 \end{figure}

 %When $\sigma=0$, i.e., $p=\mu$, we find that the above result can also be written as $H(n-1,1-2\mu)$,
 where $H$ is the harmonic function. The asymptotics of the harmonic function tell us that $L_{\text{ext}}(n) \sim n^{2\mu_p}$, therefore, if $\sigma_p=0$, the external collaboration is super-scaling. Sadly, when $\sigma_p>0$ the formula cannot be written more compactly. To make this formula numerically easier to compute, however, we can approximate this sum as an integral, which does not affect the results; instead these values are effectively the same, but now much easier to compute. Even  for $sigma_p>0$, scaling is approximately a power-law.  This theoretical curve is plotted in Fig.~\ref{fig:LextTheory}a. We show that institutions that appear earlier (e.g., $N_i=10^3$) have a smaller scaling law than those that appear later (e.g., $N_i=3 \times 10^{5}$), and finite variance in $p$ creates larger scaling laws than no variance. Because institutions grow linearly with $N_i$, this implies that smaller institutions should have a larger scaling law than larger ones \cite{Tria2014}. We can also create a histogram of the scaling exponents in Fig.~\ref{fig:LextTheory}b. Because the cumulative number of institutes grows as $N^{\nu/\rho}=N^{1/2}$, the number of new institutes scales as $N^{-1/2}$, therefore we sample exponents with this frequency. External collaborations therefore create heterogeneous scaling exponents independent of $p$. The heterogeneous scaling is instead an \emph{emergent property}.%Let $N_i$ be the value of $N$ when that institute first appears.  In this figure, we notice that, while the $\sigma_p=0$ theory produces no dependence on $N_i$, there is a surprising dependence on $N_i$ for the $\sigma_p>0$ theory. Based on the numerical relation between the scaling exponent and $N_i$, and the rate of new institutes as a function of $N$, we can create a histogram of the external collaboration exponents up to $N=10^6$, . %We notice a significant variance in these exponents, between 1 and 2, in agreement with what we find in simulations (see main text Fig. 3 and Fig.~\ref{fig:PoissonVsDeterministic}). That said, the reason for the relation is a dependence on $N_i$, which does not agree with what we find in simulations. Namely, the %final  size, $n=N/N_i$, where $N$ is fixed. This implies that the external collaboration exponent should depend on the final institution size, but that is not what we find in simulations or in data (see main text Fig. 3d). Moreover, the variance Fig.~\ref{fig:LextTheory}b is much smaller than what we find in simulations, shown in Fig.~\ref{fig:PoissonVsDeterministic}. We therefore have to conclude that this theory is too simplistic, but at least begins to explain the relevant phenomena that we see.

\section{Comparing Theory To Simulations}%Simulations of Alternative Mechanisms}

%Theory surrounding the Polya's urn portion of our model is discussed in detail in previous work \cite{Tria2014}. Nonetheless, we check the robustness of this theory in Fig.~\ref{fig:SimHeapsZipfs}. We find excellent agreement between the theory and simulation, demonstrating that, even for finite sizes, the theory they developed accurately explains the simulation patterns. This shows that our simulated institutions grow sublinearly with the number of researchers, and regardless of the parameters, we consistently see a Zipf's law in the distribution of institution sizes. 

% \begin{figure}[thb!]
%    \centering
%    \includegraphics[width=\columnwidth]{figures/SimScaling.pdf}
%    \caption{Simulation of (a) Heaps law and (b) Zipf's for various values of $\rho$ and $\nu$. Dashed lines are theoretical scaling exponents.
%    }
%    \label{fig:SimHeapsZipfs}
%\end{figure}

%\subsection*{Internal and External Collaboration Scaling}
\begin{figure}[ht]
    \centering
    \includegraphics[width=0.5\textwidth]{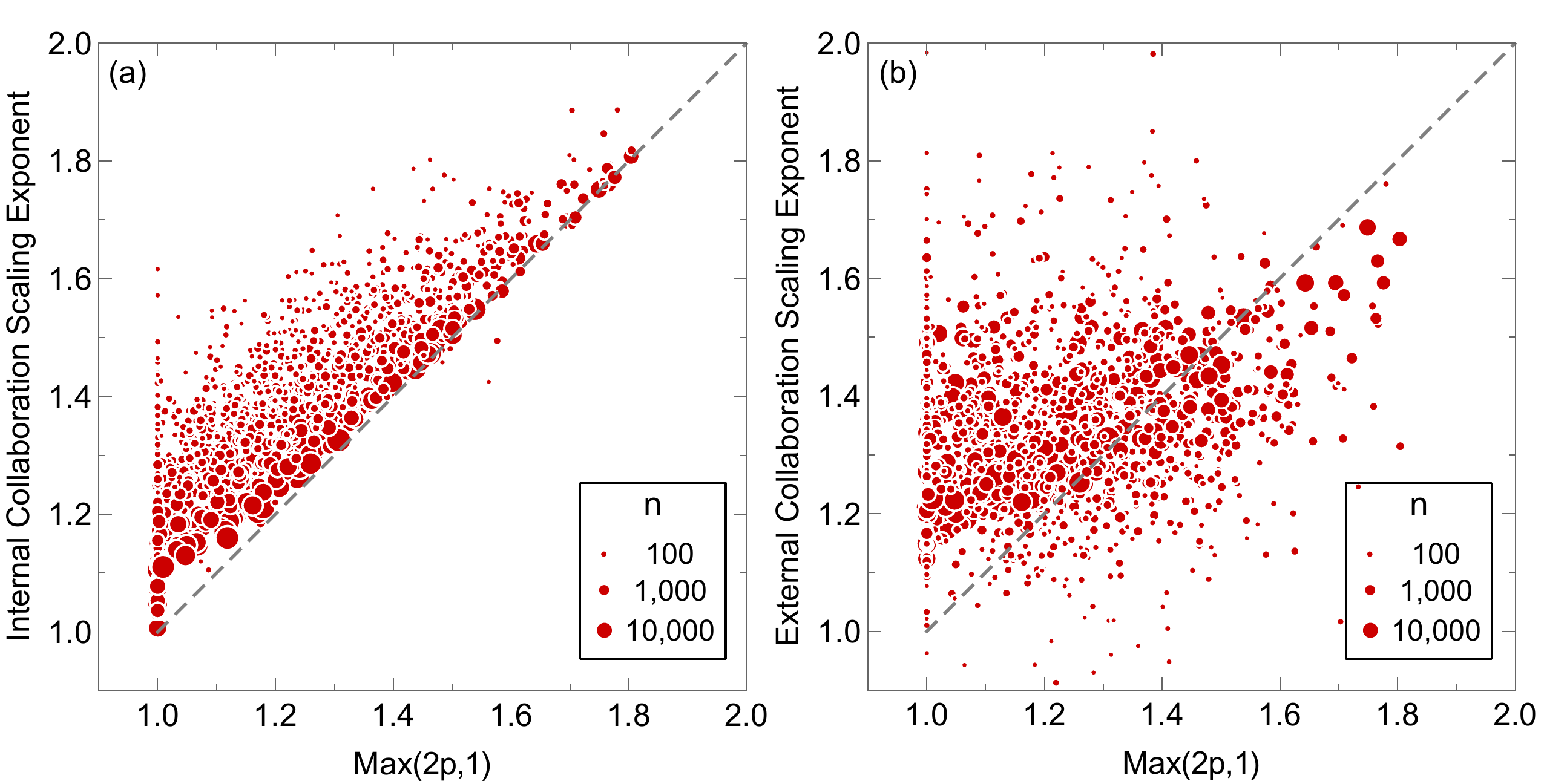}
    \caption{Simulated and theoretical collaboration scaling. (a) Internal collaborations are expected to scale as $\sim n^{2 p}$ (Eq.~\ref{eq:intscale}), which agrees well with simulations, especially for large institution sizes, $n$. (b) In contrast, the external collaborations do not strongly correlate with the internal scaling theory ($s=0.22$, p-value $<10^{-6}$). Data gathered for 15 simulations for institutions with final size, $n>50$.}
    \label{fig:PSimTheory}
\end{figure}

We first compare theory with simulations for internal collaboration scaling exponents. Equation~\ref{eq:intscale} predicts that, for $p>1/2$, $L_\text{int}\sim n^{2p}$, therefore, we should see a significant correlation between the simulation scaling laws and $2p$, especially for large $n$. Figure~\ref{fig:PSimTheory} compares the simulation and theoretical exponents for 15 simulations with $\rho=4$, $\nu=2$, $\mu_p=0.6$, and $\sigma_p=0.1$ after 500K time steps. We recorded from $n\ge 10$ for institutions with more than $50$ researchers at the final time step. In total there were 1582 simulated institutions studied. We find agreement with theory for large $n$ in Fig.~\ref{fig:PSimTheory}a, and overall a significant correlation with theory (Spearman rank correlation, $s=0.85$, p-value $<10^{-6}$). We also find good agreement with simulations in Fig.~\ref{fig:VarVsNoVarSims}a, which further demonstrates that, as expected, variance in $p$ creates variance in the scaling exponents. Moreover, we can focus on $\sigma_p=0.0$ data, and notice that, as $n$ becomes large, we have better and better agreement with the theory. The broad distribution of scaling exponents for both internal and external collaborations can be seen in the Appendix.

We next compare external collaboration with theory in Fig.~\ref{fig:PSimTheory}b. Equations~\ref{eqn:Lext0} and \ref{eqn:Nkext} predict low correlation between exponents and the $p$ parameters, which we also observe in simulations %. Instead, the scaling is dominated by the mean scaling parameter, with a sub-dominant effect due to variance in $p$. We find only a weak correlation with $p$ 
($s=0.22$, p-value $<10^{-6}$), showing support for the theory's qualitative distinction between internal and external scaling. We also see qualitative agreement with theory in Fig.~\ref{fig:VarVsNoVarSims}b. Namely, we see that $\sigma=0.0$ theory is in reasonably good agreement with simulations, even for small $n$. With $\sigma_p=0.1$, we find that scaling exponents tend to be larger, in agreement with Equations~\ref{eqn:Lext0} and \ref{eqn:Nkext}, which shows that the scaling exponents increase with $\sigma_p$. 

That being said, the theory implies that final institution size is proportional to its age, and therefore we should see a correlation between the final institution size and the scaling exponent (Fig.~\ref{fig:LextTheory}). We find, however, no significant correlation with size (p-value$=0.21$), as shown in the Appendix. Moreover, there is a significant correlation between internal and external scaling laws in simulations that is also not captured by the theory. %Finally, we recall the statistically significant correlation between external collaboration scaling and $p$. 
These findings together demonstrate that the present theory does not fully describe the dynamics. Nonetheless, we are able to describe much of the behavior, including heterogeneous external collaboration scaling. 

%{\color{red} discuss simulations with and without variance. Discuss: large variance in external scaling with p variance (agreement with theory). Internal collaboration scaling: unexpected dependence on final size (finite size effect).}
%{\color{blue} Figure~\ref{fig:PSimTheory} demonstrates comparisons between simulations and theory. While internal collaboration scaling matches well with theory, external collaboration depends only weakly on $p$, in qualitative agreement with Eq.~\ref{}. Moreover, there is little relation between institution size and the scaling parameter, in contrast with Eq.~\ref{}.}

\section{Conclusion}

Burghardt et al. \citeyear{Burghardt2020} found surprising statistical regularities in the growth of research institutions, and created a model to explain these regularities. We explore this model in greater detail and discover empirical agreement with model assumptions and realistic network properties such as significant community structure. Furthermore, we produce a theoretical grounding for this model and show agreement between theory and simulations. This theory demonstrates that while the the internal collaboration exponent is proportional to $p$, the external collaboration scaling parameter is approximately independent of all other parameters. %Finally, we show that cross-sectional scaling varies in time, in agreement with Burgardt et al., and that there are correlations between internal and external scaling laws that are captured by our model.

While these findings ground the Burghardt et al.'s model in a stronger empirical and theoretical foundation, there are limitations in what we can explain. First, the growth of institutions is sub-linearly related to its size ($\Delta n\sim n^{0.9}$), while the model predicts a linear relation. Second, while collaborations often form between friends of friends, this is not always the case, as the geodesic mean distance in Fig.~\ref{fig:Densify} is greater than two. Finally, the model does not fully explain how the institution size correlates to the total number of researchers hired (Fig.~\ref{fig:InstSizeVsHires}). While it is expected to be linear, we see significant deviations, either due to the simplicity of our model, or potentially limitations in data collected prior to 1950 \cite{Burghardt2020}. While agreement between model and data is still pretty close, these deviations suggests limitations of the current model to fully describe data. In addition, the theory predicts a much higher value for $\langle k_\text{ext}\rangle$ than we see in simulations, shown in Fig.~\ref{fig:KMean}. Similarly, the external collaboration distribution for simulations seen in Fig.~\ref{fig:LextTheory} is not in agreement with Fig.~\ref{fig:PoissonVsDeterministic}, and it suggests a dependence on time. While we have made great progress, future work is needed to fully understand this model. 

\section{Appendix}

%\begin{figure*}
%    \centering
%    \includegraphics[width=0.8\textwidth]{figures/LogLikelihoodFits.pdf}
%    \caption{Normalized log-likelihood of mixed effect models of mean institution impact. We compare the quality of the model fit as we add features to the model. Features are added in a greedy manner, where each additional feature increases the log-likelihood of the mixed model of (a) computer science, (b) physics, (c) math, and (d) sociology the most.}
%    \label{fig:allmodelfits}
%\end{figure*}

\begin{figure}[b!]
    \centering
    \includegraphics[width=0.5\textwidth]{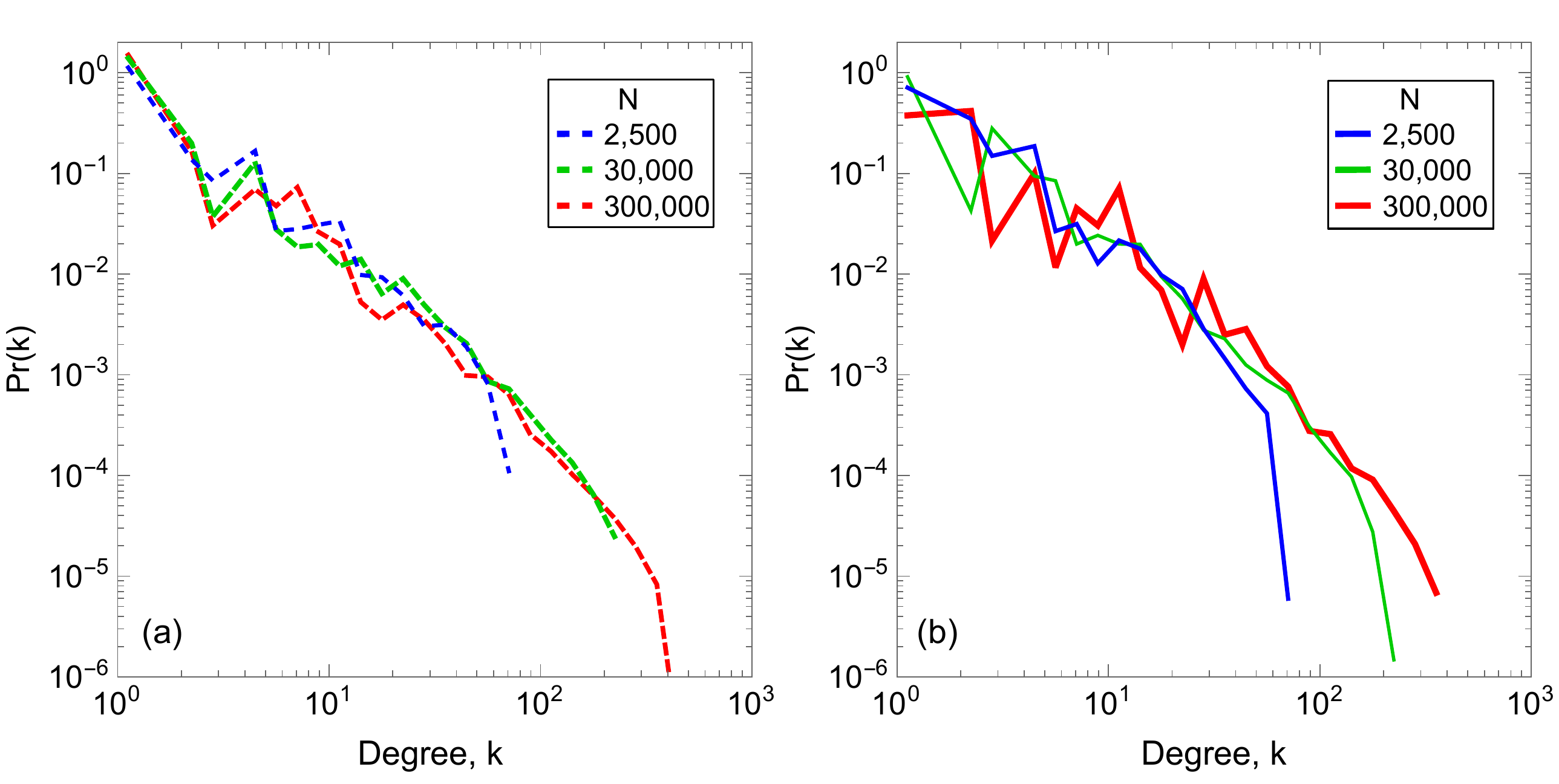}
    \caption{Degree distribution versus network size for $\sigma_p=0.0$ and 0.1.}
    \label{fig:DegreeDist}
\end{figure}

We first check the robustness of the heavy-tailed degree distribution versus network size and $\sigma_p$, which is shown in Fig.~\ref{fig:DegreeDist}. In this data, which is averaged over 10 simulations, we find the wide degree distribution is robust to variations in model size and $\sigma_p$. %Values of $\mu_p$, $\rho$, and $\nu$ similarly do not strongly affect results.

\begin{figure}[ht]
    \centering
    \includegraphics[width=0.5\textwidth]{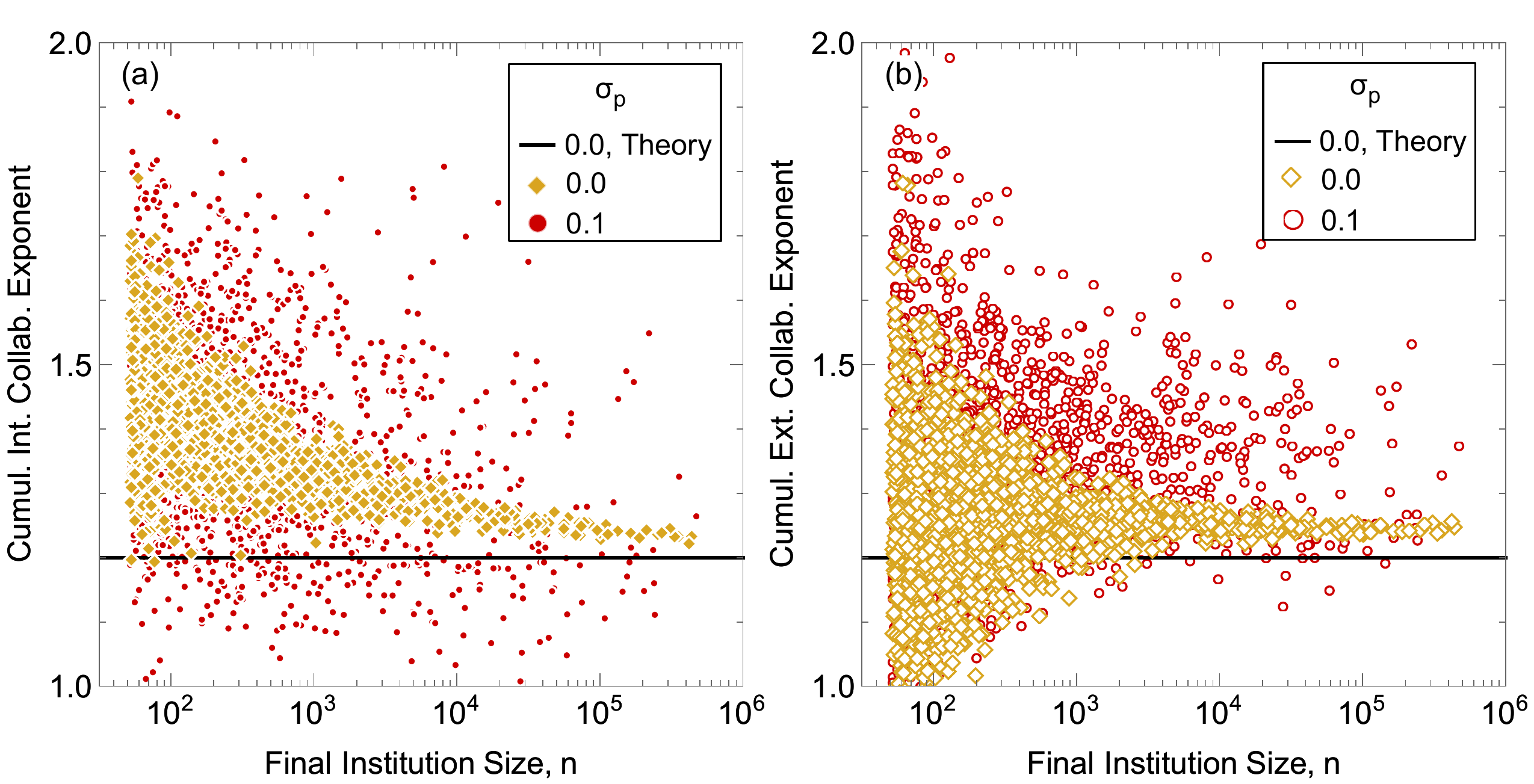}
    \caption{Simulations with different scaling laws ($\sigma_p=0.1$) and constant collaboration scaling laws ($\sigma_p=0.0$). (a) Internal collaboration scaling exponents versus %final 
    institution size, and (b) external collaboration scaling exponents versus institution size. In all simulations, $\rho=4$, $\nu=2$, and $\mu_p=0.6$. Simulations are run 15 times, with 1.6 K institutions gathered for each parameter such that $n>50$ and scaling laws are taken for $n\ge 10$.}
    \label{fig:VarVsNoVarSims}
\end{figure}

Next, we compare the collaboration scaling exponents for simulations with $\sigma_p=0.0$ and $0.1$ in Fig.~\ref{fig:VarVsNoVarSims}. We find that, due to stochasticity, both parameters show variance in the collaboration exponents, although when $\sigma_p=0.0$, the variance decreases significantly with $n$, while for $\sigma_p=0.1$, the variance is high even for $n=10^5$.

%\begin{figure}[ht]
%    \centering
%    \includegraphics[width=0.5\textwidth]{figures/FigS22.pdf}
%    \caption{Internal and external cross-sectional collaboration exponents for alternative simulation models. (a) Internal and external exponents versus the cumulative number of researchers for simulations with $\lambda=1$ Poisson distributed numbers of initial collaborators (on average one internal collaborator, and one external collaborator). (b) The same figure for the current simulation with exactly one internal and one external collaborator.}
%    \label{fig:PoissonVsDeterministic}
%\end{figure}

\begin{figure}[h!]
    \centering
    \includegraphics[width=\linewidth]{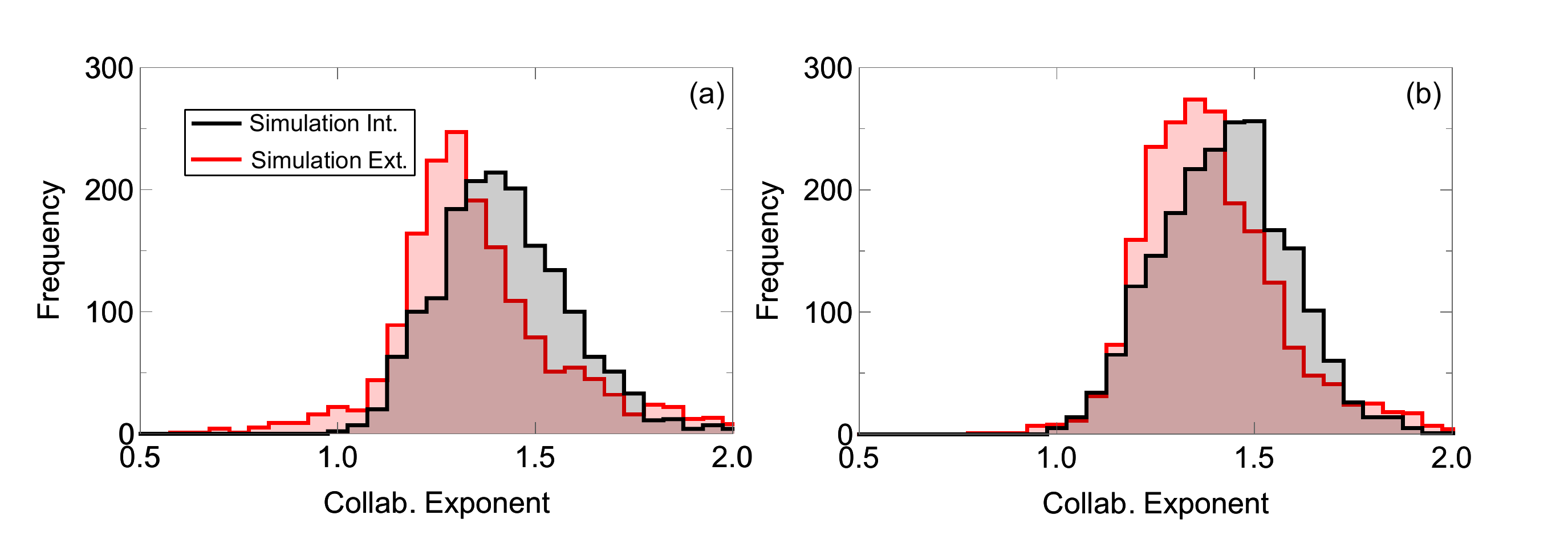}
    \caption{Internal and external longitudinal collaboration exponents for alternative simulation models. (a) Internal and external exponents for simulations with $\lambda=1$ Poisson distributed numbers of initial collaborators (on average one internal collaborator, and one external collaborator). (b) The same histograms for the current simulation with exactly one internal and one external collaborator.}
    \label{fig:PoissonVsDeterministic}
\end{figure}

Finally, we explore robustness of results with respect to realistic model variants. The theoretical analysis predicts heterogeneous densification of collaborations within institutions and  heterogeneous densification  for external collborations (between institutions). We check each of these qualitative results in Fig.~\ref{fig:PoissonVsDeterministic}, and compare these results for two variants of our model to check sensitivity of our analysis. In the original model, hired ``researchers'' initially create one internal and one external link. We look at a variant of this simulation in which the number of initial internal and external collaborators was Poisson distributed, with $\lambda=1$ (i.e., on average one internal and one external collaborator).  Importantly, Bhat et al. and Lambiotte et al. shows that number of links over time are not self-averaging in their densification model \cite{Lambiotte2016,Bhat2016}, therefore initial conditions greatly affect the final number of links and may affect the observed scaling behavior. Figure~\ref{fig:PoissonVsDeterministic} shows our results. We find that, while there are slightly more outliers in the scaling exponent distribution, results are quantitatively very similar. Overall this suggests that our model robustly creates agreement their theory. 

\end{document}